\documentclass[twocolumn]{article}
\usepackage{parskip}
\usepackage{amssymb}
\usepackage{amsmath}
\usepackage{fullpage}
\usepackage{authblk}
\usepackage{blindtext}
\usepackage{siunitx}
\usepackage{float}
\usepackage{caption}
\usepackage[superscript,nomove]{cite}
\usepackage{hyperref}
\usepackage[normalem]{ulem}
\usepackage{graphicx}
\usepackage{cancel}
\usepackage{algpseudocode}
\usepackage{cprotect}
\usepackage{xcolor}
\usepackage{circuitikz}
\usepackage{epstopdf}

\makeatletter \renewcommand{\@citess}[1]{\textsuperscript{[#1]}} \makeatother

\makeatletter
\def\@firstoftwo@second#1#2{%
  \def\temp##1.##2\@nil{##2}%
   \temp#1\@nil}
\newcommand\sref[1]{%
   (A.\expandafter\@setref\csname r@#1\endcsname\@firstoftwo@second{#1})%
}

\makeatother

\title{Diffusion in a $d$-dimensional rough potential}
\author[1]{\small Jacob Jeffries \thanks{jwjeffr@g.clemson.edu}}
\author[2]{\small Emilio Mendoza Reyes}
\author[3]{\small Fadi Abdeljawad}
\author[4]{\small Murray Daw}
\author[1,5]{Enrique Martinez \thanks{enrique@clemson.edu}}
\affil[1]{Department of Materials Science and Engineering, Clemson University, Clemson, SC 29634, USA}
\affil[2]{School of Computing, Clemson University, Clemson, SC 29634, USA}
\affil[3]{Department of Materials Science and Engineering, Lehigh University, Bethlehem, PA 18015, USA}
\affil[4]{Department of Physics and Astronomy, Clemson University, Clemson, SC 29634, USA}
\affil[5]{Department of Mechanical Engineering, Clemson University, Clemson, SC 29634, USA}
\date{\small \today}

\renewenvironment{abstract}
 {\quotation\small\noindent\rule{\linewidth}{.5pt}\par\smallskip
  {\centering\bfseries\abstractname\par}\medskip}
 {\par\noindent\rule{\linewidth}{.5pt}\endquotation}

\begin{document}

\twocolumn[
  \begin{@twocolumnfalse}
  \maketitle
    \begin{abstract}
        The prediction of diffusion in solids is necessary to understand the microstructure evolution in materials out of equilibrium. Although one can reasonably predict diffusive transport coefficients using atomistic methods, these approaches can be very computationally expensive. In this work, we develop an analytical model for the diffusivity in a noisy solid solution in an arbitrary number of dimensions ($d$) using a mean first passage time analysis. We observe that roughness always decreases the diffusivity, aligning with sluggish diffusion theories in concentrated alloys, finding that an increase in diffusivity induced by alloying elements must be due to a decrease in the average activation energy, not to the noise. These analytical results are then compared with kinetic Monte Carlo simulations, which are in good quantitative agreement with the simulation data for $d\leq 5$, and excellent quantitative agreement for $d\leq 3$. This generalization to arbitrary dimensions has been elusive to the community since Zwanzig \cite{Zwanzig1988} published his seminal work on 1-dimensional systems.
    \end{abstract}
  \vspace{0.5cm}
  \end{@twocolumnfalse}
]


Accurately modeling diffusion in solids is paramount to fundamentally understand the microstructure evolution in materials out of equilibrium. The calculation of transport coefficients in solids is necessary for modeling radiation-induced segregation\cite{wiedersich1979theory}\cite{nastar20121}\cite{rezwan2022effect}\cite{PIOCHAUD2016249}, grain boundary diffusion\cite{CAHN19974397}, solid-state sintering\cite{WANG2006953}, pitting corrosion \cite{SCHEINER20092898}\cite{jafarzadeh2019computational}, and more. In a homogeneous pure solid, transport coefficients, such as the diffusivity $D$, are often modeled using harmonic transition state theory (HTST)\cite{htst}:

\begin{equation}
    D = D_0 \exp\left(-\frac{\Delta E}{k_BT}\right)
\end{equation}

where $D_0$ is the rate prefactor, which can be estimated from the vibrational modes at the initial and transition states\cite{VINEYARD1957121}\cite{10.1063/1.5086746} and a hopping length, $k_B$ is the Boltzmann constant, $T$ the absolute temperature, and $\Delta E$ the activation energy of diffusion, which can be estimated using the nudged elastic band (NEB) method\cite{10.1063/1.1323224}\cite{10.1063/1.1329672}\cite{Nakano2008ASP}\cite{MARAS201613}\cite{10.1063/1.1627754} coupled with atomistic simulation softwares \cite{LAMMPS}\cite{Case2023}\cite{Brooks2009-kz}\cite{PhysRevB.47.558}\cite{KRESSE199615}\cite{PhysRevB.54.11169}.

However, for multi-component solid solutions, the activation energy for a diffusive hop depends on local chemistry, varying across the system. The effect of noisy energy barriers has been well-studied in 1 dimension ($1d$). For example, Zwanzig derived an expression in 1d for diffusion in a solid with roughness $\varepsilon^2$ and barrier mean $\mu$\cite{Zwanzig1988}:

\begin{equation}
    D = D^*\exp\left(-\left(\frac{\varepsilon}{k_BT}\right)^2\right)
\end{equation}

where $D^*$ denotes the diffusivity of the walker in a single-element homogeneous solid with barrier $\mu$:

\begin{equation}\label{eq:homogeneous}
    D^* = D_0 \exp\left(-\frac{\mu}{k_BT}\right)
\end{equation}

This expression shows that, in $1d$, the addition of noise in a solid solution with equal mean as the pure system suppresses diffusivity, which relates to the controversial sluggish diffusion effect in high entropy alloys\cite{tsai2013sluggish}\cite{divinski2018mystery}\cite{dkabrowa2019demystifying}. However, a similar general analytical expression in a higher number of dimensions has been elusive to the community for over four decades. To estimate diffusivities in higher dimensions, one can perform atomistic simulations such as atomistic kinetic Monte Carlo (KMC) \cite{SENNINGER20161}. However, these simulations can be expensive, highlighting the need for an analytical or semi-analytical expression instead for faster estimates and deeper physical insights.

Seki and Bagchi derived such an expression, concluding that $D = D^*\exp\left(-\frac{1}{2}\left(\frac{\varepsilon}{k_BT}\right)^2\right)$ for all $d\geq 2$ \cite{seki2015relationship}. We find in this work, however, that such an expression is valid for $d = 2$, but not $d > 2$. Seki et al. later derived an expression that is consistent with Monte Carlo results for $d = 2$ and $d = 3$, but is inconsistent with Zwanzig's expression for $d = 1$, concluding that diffusion in $1d$ is pathologically different to higher dimensions \cite{seki2016anomalous}.

Such an analytical expression, i.e. one that predicts diffusivity for all dimensionalities, can expedite the analysis of diffusion in dimensionality-dependent phenomena, such as one-dimensional diffusion along dislocations\cite{doi:10.1080/01418617908239293}\cite{legros2008observation} or DNA\cite{wang2006single}\cite{gorman2008visualizing}\cite{gorman2010visualizing}, two-dimensional diffusion along a grain boundary\cite{peterson1983grain}\cite{mishin1997grain}, or three-dimensional diffusion in bulk\cite{vaidya2016ni}\cite{beke2016diffusion}\cite{vaidya2018bulk}. Additionally, an analytical expression would help clarify the controversy behind the sluggish diffusion effect. In this work, we present an analytical expression for diffusivity in $d$ dimensions on a lattice with normally distributed energy barriers with mean $\mu$ and variance $2\sigma^2$ that is consistent with Zwanzig's expression for $d\leq 5$, with excellent quantitative agreement for $d\leq 3$ and $\beta\sigma \leq 2$.


The hopping rate over an energy barrier $\Delta E$ is estimated by HTST:

\begin{equation}
    R = \nu \exp\left(-\beta \Delta E\right)
\end{equation}

where $\nu$ is the rate prefactor and $\beta = 1/k_BT$.

\begin{figure}[H]
    \centering
    \includegraphics[width=\linewidth]{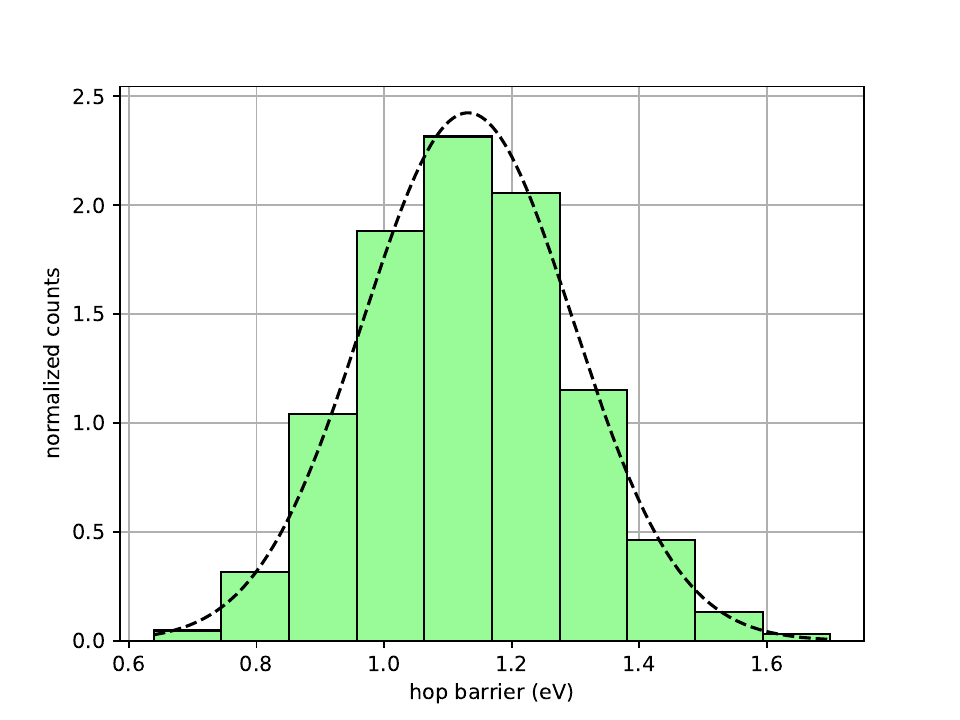}
    \caption{Example migration barrier distribution for a vacancy hop in a model $\text{Fe}_{30}\text{Ni}_{30}\text{Cr}_{40}$ solid solution using the climbing image NEB method in LAMMPS with an EAM potential, fit to a normal distribution.}
    \label{fig:barrier-dist}
\end{figure}

In a homogeneous single-component solid, there is a singular lattice point energy $E^\circ$ and a single saddle point energy $E^\dagger$, yielding an energy barrier $\Delta E = E^\dagger - E^\circ$. In a heterogeneous multi-component solid, however, there are multiple lattice and saddle point energies, characterized by local chemistries. For many random solid solutions, the migration barriers can be modeled as a normal random variable. As an example, we have computed the distribution of vacancy-mediated migration barriers in an $\text{Fe}_{30}\text{Ni}_{30}\text{Cr}_{40}$ model alloy using the climbing image NEB method in LAMMPS using an embedded atom method potential \cite{ternary-eam} by Zhou et al. (Figure \ref{fig:barrier-dist}), which is relatively well-modeled by a normal distribution.

A closely related approach was recently developed by Xu et al., who combined machine-learning-assisted KMC with a species-resolved analytical diffusion model for vacancy-mediated self-diffusion in multi-principal element alloys \cite{xu2023mechanism}. Their model explicitly separates random-barrier and random-trap contributions, incorporates species-dependent migration-barrier and site-energy distributions, and accounts for vacancy correlation effects. This allowed them to show that sluggish diffusion is not controlled by the width of the migration-barrier distribution alone; instead, site-energy variations, chemically biased jump probabilities, and vacancy back-correlation can be essential. The present work addresses a complementary limiting problem: rather than constructing a species-resolved model for a particular alloy chemistry, we derive a compact dimension-dependent expression for a dilute walker in an idealized statistically rough landscape. This isolates the role of dimensionality and potential roughness, while deliberately neglecting species-resolved correlations and trap/barrier coupling beyond the assumed distributions of initial and saddle energies.

\begin{figure}[H]
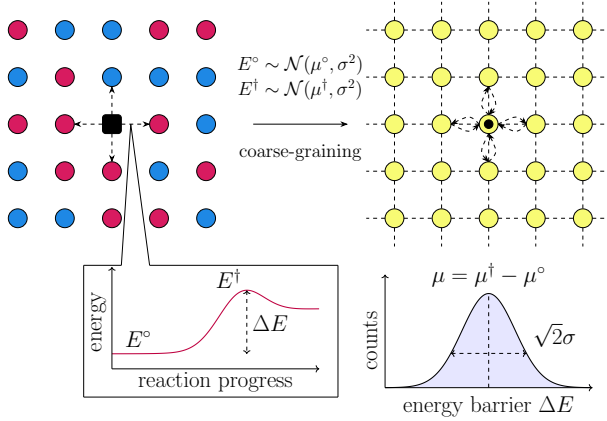

    \centering
    \include{plots/tikz-circuit-solution}
    \caption{Coarse-grained migration barriers in an example binary solid solution with a vacancy.}
    \label{fig:coarse-grain}
\end{figure}

As such, we begin our model by coarse-graining local chemistries into two distributions $E^\dagger$ and $E^\circ$ (Figure \ref{fig:coarse-grain}), which are respectively the energies at the saddle and initial points. This reduction to a single barrier is an intentional coarse-graining of the full reaction-coordinate landscape, whose detailed shape can itself influence mean first-passage times when intermediate barriers or structured potential profiles are present \cite{kumar2025speeding}. We then assume that $E^\dagger$ and $E^\circ$ can be modeled as independently distributed random variables:

\begin{equation}\label{eq:e-normal-dist}
    \begin{aligned}
        E^\circ &\sim \mathcal{N}\left(\mu^\circ, \sigma^2\right) \\
        E^\dagger &\sim \mathcal{N}\left(\mu^\dagger, \sigma^2\right)
    \end{aligned}
\end{equation}

where $\mathcal{N}(\mu, \sigma^2)$ denotes a normal distribution with mean $\mu$ and variance $\sigma^2$. The resulting diffusive energy barriers are then similarly normally distributed:

\begin{equation}
    \Delta E = E^\dagger - E^\circ \sim \mathcal{N}(\mu, 2\sigma^2)
\end{equation}

where we have defined the mean energy barrier $\mu = \mu^\dagger - \mu^\circ$ for brevity.

The hopping rates in the solution are then:

\begin{equation}\label{eq:e-rates}
    \lambda = \nu \exp\left(-\beta\Delta E\right) \sim \nu \mathcal{L}\left(-\beta \mu, 2\beta^2\sigma^2\right)
\end{equation}

where $\mathcal{L}$ denotes a lognormal distribution, and $\nu$ is implicitly assumed to be constant throughout the solution, and independent of temperature.

In a single lattice site in a solid solution, a given walker will see $n$ transitions, all with transition rates drawn from the distribution of $\lambda$. Before constructing the \(d\)-dimensional escape problem, we first introduce a
trajectory-level coarse-graining of the rough potential along a representative
directed transport coordinate. Along such a directed path, successive barriers are
encountered in series. For a diffusion path with transition times $(t_1, t_2, \cdots, t_N)$, the total transition time is $t_\text{tot} = t_1+t_2+\cdots+t_N$. This yields a characteristic rate $R_\text{eff}$:

\begin{equation}\label{eq:harmonic-addition}
    \begin{aligned}
        R_\text{eff} &= \frac{1}{t_\text{tot}} = \frac{1}{1/r_1 + 1/r_2 + \cdots + 1/r_N} \\
        &= \frac{1}{N}\left(\frac{1}{N}\sum_{\ell=1}^N \frac{1}{r_\ell}\right)^{-1} = \frac{1}{N}\left\langle r_\ell^{-1}\right\rangle^{-1}
    \end{aligned}
\end{equation}

where $r_i = 1/t_i$ is the characteristic rate of transition $i$. Then, in the $N\to\infty$ limit, the characteristic rate along the path $r_\text{eff} = NR_\text{eff}$ is exactly the harmonic mean of $\lambda$:

\begin{equation}\label{eq:effective-rate}
    r_\text{eff} = \left\langle \lambda^{-1}\right\rangle^{-1} = \nu\exp(-\beta\mu-\beta^2\sigma^2)
\end{equation}

where we have used the result that the harmonic mean of a lognormal distribution $\mathcal{L}(\mu, \sigma^2)$ is $\exp(\mu-\sigma^2/2)$. This result is consistent with Zwanzig's result, i.e. that the diffusivity in a noisy potential in one dimension is $D\propto \exp(-\beta\mu - \beta^2\sigma^2)$ \cite{Zwanzig1988}.

Using this characteristic rate, we describe diffusion paths in the noisy potential as single hops with an effective barrier $\mu_\text{eff}$:

\begin{equation}\label{eq:effective-barrier}
    \beta\mu_\text{eff} = \beta\mu + \beta^2\sigma^2
\end{equation}

which yields an effective hopping rate $\lambda'$ for one jump:

\begin{equation}
    \begin{aligned}
        \lambda' &\sim \nu\mathcal{L}(-\beta\mu_\text{eff}, 2\beta^2\sigma^2) \\
        &= \nu\exp\left(-\beta^2\sigma^2\right)\mathcal{L}(-\beta\mu, 2\beta^2\sigma^2)
    \end{aligned}
\end{equation}

From our coarse-grained description, we return to the \(d\)-dimensional problem. At a coarse-grained point in the effective landscape, the walker has \(n=2d\) possible outgoing effective directional transitions. The corresponding passage times are then exponentially distributed, where $i$ indexes possible transitions:

\begin{equation}
    T_i'\;|\;\lambda_i \sim \text{Exp}\left(\lambda_i'\right) = \exp\left(\beta^2\sigma^2\right)\text{Exp}\left(\lambda_i\right)
\end{equation}

The escape time from this coarse-grained point is determined by the standard
KMC total-rate rule, i.e. we want to sample the first passage time for the walker conditioned on the realization of the rate parameters:

\begin{equation}
    \begin{aligned}
        T'\; | \; \lambda_1, \cdots, \lambda_n &= \min_{1\leq i\leq n}T_i'\;|\;\lambda_i\\
        &\sim \exp(\beta^2\sigma^2)\min_{1\leq i\leq n}\text{Exp}\left(\lambda_i\right)\\
        &\sim \exp(\beta^2\sigma^2)\;\text{Exp}\left(\sum_{i=1}^n \lambda_i\right)
    \end{aligned}
\end{equation}

We can then find $T'$ via the law of total expectation:

\begin{equation}
    \begin{aligned}
        T' &= \mathbb{E}_{\lambda_1, \cdots, \lambda_n}\left[  T'\; | \; \lambda_1, \cdots, \lambda_n \right]\\
        &=\exp(\beta^2\sigma^2)\left(\sum_{i=1}^n \lambda_i\right)^{-1}
    \end{aligned}
\end{equation}

We are then concerned with computing the mean first passage time $\mathbb{E}[T']$. From this expectation, we can then evaluate the ratio of the MFPT and the noise-less MFPT:

\begin{equation}\label{eq:ratio}
    \begin{aligned}
        \phi &= \frac{\mathbb{E}[T']} {\lim_{\beta^2\sigma^2\to 0}\mathbb{E}[T']} = n\nu \exp\left(-\beta\mu\right)\mathbb{E}[T'] \\
        &=n\nu \exp\left(-\beta\mu+\beta^2\sigma^2\right)\mathbb{E}\left[\frac{1}{\sum_{i=1}^n \lambda_i}\right] \\
        &=n \exp(\beta^2\sigma^2)\mathbb{E}\left[\frac{1}{\sum_{i=1}^n \exp(V_i)}\right]
    \end{aligned}
\end{equation}

where $V_i\sim\mathcal{N}(0, 2\beta^2\sigma^2)$, and where we have used the fact that $V_i$ is equivalent to $-V_i$ in the distributional sense. We can approximate the random variable $\frac{1}{\sum_{i=1}^n e^{V_i}}$ by first defining the auxiliary random variable $\delta_i$:

\begin{equation}
    \exp(V_i) = 1 + \delta_i
\end{equation}

as well as the sum of these variables $S = \sum_{i=1}^n \delta_i$. Then:

\begin{equation}
    \frac{1}{\sum_{i=1}^n \exp(V_i)} = \frac{1}{n+S} = \frac{1}{n}\sum_{k=0}^\infty (-1)^k \frac{S^k}{n^k}
\end{equation}

Therefore:

\begin{equation}
    \phi = \exp(\beta^2\sigma^2)\sum_{k=0}^\infty (-1)^k \frac{\mathbb{E}\left[S^k\right]}{n^k}
\end{equation}

where $\mathbb{E}[S^k]$ can be evaluated using the multinomial theorem:

\begin{equation}
    \begin{aligned}
        \mathbb{E}[S^k] &= \mathbb{E}\left[\left(\sum_{i=1}^n \delta_i\right)^k\right] \\
        &= \sum_{\substack{m_1,\cdots,m_n\geq 0 \\ m_1+\cdots+m_n = k}}\frac{k!}{m_1!\cdots m_n!}\mathbb{E}\left[\prod_{i=1}^n \delta_{m_i}\right]
    \end{aligned}
\end{equation}

To first order, we approximate each barrier within the coarse-grained effective potential as independent of one another. Therefore:

\begin{equation}
    \begin{aligned}
        \mathbb{E}[S^k] = \sum_{\substack{m_1,\cdots,m_n\geq 0 \\ m_1+\cdots+m_n = k}}\frac{k!}{m_1!\cdots m_n!}\prod_{i=1}^n \mu_{m_i}
    \end{aligned}
\end{equation}

where $\mu_{m_i} = \mathbb{E}[\delta_i^m]$ is the $m$'th moment of $\delta_i$. For $m\geq 0$:

\newcommand{\stirling}[2]{\genfrac{\{}{\}}{0pt}{}{#1}{#2}}

\begin{equation}
    \begin{aligned}
        \mu_m &= \mathbb{E}[\left(\exp\left(V_i\right) - 1)^m\right]\\
        &= \sum_{j=0}^m\binom{m}{j}(-1)^{m-j}\mathbb{E}\left[\exp\left(j V_i\right)\right]\\
        &= \sum_{j=0}^m\binom{m}{j}(-1)^{m-j} \exp(j^2\beta^2\sigma^2) \\
        &=m!\sum_{r=0}^\infty \stirling{2r}{m}\frac{\beta^{2r}\sigma^{2r}}{r!}
    \end{aligned}
\end{equation}

where $\stirling{q}{m}$ is the Stirling number of the second kind:

\begin{equation}
    \stirling{q}{m} = \frac{1}{m!}\sum_{j=0}^m (-1)^{m-j}\binom{m}{j}j^q
\end{equation}

Here, we make the approximation that $\beta\sigma$ is small, and truncate to order $\mathcal{O}(\beta^4\sigma^4)$. Then:

\begin{equation}
    \mu_m = m!\left( \stirling{0}{m} + \stirling{2}{m}\beta^{2}\sigma^{2}\right) + \mathcal{O}(\beta^4\sigma^4)
\end{equation}

To this order, very few terms are non-zero. Namely, $\stirling{0}{m} = 0$ for $m \geq 1$ and $\stirling{2}{m} = 0$ for $m \geq 3$. Therefore, after truncating away any $\mathcal{O}(\beta^4\sigma^4)$ terms, the only non-zero moments $\mu_m$ are:

\begin{equation}
    \begin{aligned}
        \mu_0 &= 1 \\ 
        \mu_1 &= \beta^{2}\sigma^{2} + \mathcal{O}(\beta^4\sigma^4) \\
        \mu_2 &= 2\beta^{2}\sigma^{2} + \mathcal{O}(\beta^4\sigma^4) \\
    \end{aligned}
\end{equation}

Notably, the expression for $\mathbb{E}[S^k]$ is a polynomial in $\mu_m$. Truncating up to order $\mathcal{O}(\beta^4\sigma^4)$ then is equivalent to only including the terms in $\mathbb{E}[S^k]$ that are linear in $(\mu_m)$, i.e.:

\begin{equation}
    \mathbb{E}[S^k] = \begin{cases}
        1 & k = 0 \\
        n\mu_k + \mathcal{O}\left(\beta^4\sigma^4\right) & k > 0
    \end{cases}
\end{equation}

Therefore:

\begin{equation}
    \begin{aligned}
        \phi &= \exp(\beta^2\sigma^2)\left(1 - \beta^2\sigma^2 + \frac{2\beta^2\sigma^2}{n} + \mathcal{O}(\beta^4\sigma^4)\right)\\
        &=\exp(\beta^2\sigma^2)\exp\left(- \beta^2\sigma^2 + \frac{2\beta^2\sigma^2}{n} + \mathcal{O}(\beta^4\sigma^4)\right)\\
        &=\exp\left(\frac{2\beta^2\sigma^2}{n} + \mathcal{O}(\beta^4\sigma^4)\right)
    \end{aligned}
\end{equation}

To connect the effective first-passage time to a diffusivity, we now make an effective random-walk approximation. This step should be distinguished from an exact
homogenization of the original quenched disordered lattice. In the auxiliary
coarse-grained process, hops are treated as renewal events with fixed jump
length \(a\), isotropic jump directions, finite mean waiting time
\(\langle T'\rangle\), and no additional directional memory. Within this approximation, $D$ scales inversely with the first reaction time, i.e. $D \propto 1/\mathbb{E}[T]$: 

\begin{equation}\label{eq:ratio}
    \frac{D}{D^*} = \frac{1}{\phi} = \exp\left(-\frac{2\beta^2\sigma^2}{n} + \mathcal{O}(\beta^4\sigma^4)\right)
\end{equation}

where $D^*$ is the diffusivity in a noiseless medium, defined in Eq.~\eqref{eq:homogeneous}. Our results also indicate that the diffusivity strongly depends on dimensionality $d$. Notably, for $n = 2d = 4$ in a simple cubic lattice, we recover the result of Seki and Bagchi \cite{seki2015relationship} in the small $\beta\sigma$ limit, i.e. that $D/D^* = \exp(-\beta^2\sigma^2/2)$ \cite{seki2015relationship}. Lastly, for $n = 2d = 2$, our expression is consistent with that of Zwanzig's, namely $D/D^* = \exp(-\beta^2\sigma^2)$ \cite{Zwanzig1988}, and is directly comparable with that of Zwanzig's: namely that the diffusivity of the walker with $n$ possible transitions can be expressed by plugging an effective noise parameter $\sigma_\text{eff} = \sigma / \sqrt{n/2}$ into Zwanzig's expression in $1d$, where $n = 2$ along the $1d$ chain. Importantly, however, the present approximation should be understood as a two-step effective-medium construction: first, serial roughness along a representative directed trajectory is absorbed into an effective directional barrier through a harmonic mean; second, the $n=2d$ outgoing effective directions compete through the usual KMC total-rate rule. This distinction is essential, since the harmonic mean is not a substitute for the microscopic KMC residence time at a lattice node.


To validate the approach above, we have calculated the diffusivity in a noisy solution numerically with mean barrier $\mu$ and variance $2\sigma^2$. We initialize an $N_1\times N_2\times\cdots\times N_d$ simple cubic lattice in $d$ dimensions. We then identify each nearest neighbor to point $\mathbf{n}_i = \left(n_i^{(1)}, n_i^{(2)}, \cdots, n_i^{(d)}\right)$ by adding or subtracting $1$ to each component of $\mathbf{n}_i$. To account for periodic boundary conditions, this summation is done modulo $N_d - 1$ in each dimension $d$.

Then, we independently sample the stable energies at each lattice site $i$ from a normal distribution:

\begin{equation}
    E_i^\circ \sim \mathcal{N}(0, \sigma^2)
\end{equation}

For each nearest neighbor transition $i\to j$, we independently sample the energy at the saddle point from a normal distribution:

\begin{equation}
    E_{i\to j}^\dagger \sim \mathcal{N}(\mu, \sigma^2)
\end{equation}

The energy barrier is then:

\begin{equation}
    \Delta E_{i\to j} = E_{i\to j}^\dagger - E_i^\circ \sim \mathcal{N}(\mu, 2\sigma^2)
\end{equation}

which yields a transition matrix:

\begin{equation}
    R_{i\to j} = \nu \exp\left(-\beta \Delta E_{i\to j}\right)
\end{equation}

for a given temperature $T$. Then, we perform KMC \cite{BORTZ197510}
to obtain the walker's position as a function of time, $\mathbf{r}(t) = a\mathbf{n}(t)$, where $\mathbf{n}(t)$ is an interpolation of the mapping $(t_n\mapsto \mathbf{n}_k)$ and $a$ is the lattice parameter. Then, using the Einstein relation\cite{RevModPhys.15.1} we obtain the diffusivity:

\begin{equation}\label{eq:einstein-relation}
    D = \frac{1}{2d}\lim_{t\to\infty}\frac{\left\langle \|\Delta\mathbf{r}(t)\|^2\right\rangle}{t}
\end{equation}

where $\Delta\mathbf{r}(t) = \mathbf{r}(t) - \mathbf{r}(0)$ is the displacement of the particle at time $t$ and $\langle\cdot\rangle$ denotes an ensemble average. Note that, to account for the imposed periodic boundary conditions, the coordinates $\mathbf{r}(t)$ are unwrapped before being used in equation \eqref{eq:einstein-relation}. We have described the KMC simulations performed in more detail in the Supplementary Material.

Using the simulation parameters in Table \ref{tab:simulation-params}, we performed $10$ independent simulations per $\sigma$, varying $\sigma$ from $\SI{0.0}{eV}$ to $\SI{0.15}{eV}$, additionally varying the dimensionality $d$ using box sizes in Table \ref{tab:box-sizes}.

\begin{table}[H]
    \centering
    \begin{tabular}{|c|c|}
    \hline
    Variable & Value \\
    \hline
    Mean barrier $\mu$ & $\SI{0.7}{eV}$ \\
    \hline
    Temperature $T$ & $\SI{1000}{K}$ \\
    \hline
    Lattice parameter $a$ & $\SI{1}{\angstrom}$ \\
    \hline 
    Rate prefactor $\nu$ & $\SI{10}{THz}$ \\
    \hline
    Number of steps $N_\text{steps}$ & $50$ million \\
    \hline
    \end{tabular}
    \caption{Parameters used in KMC simulations}
    \label{tab:simulation-params}
\end{table}

\begin{table}[H]
    \centering
    \begin{tabular}{|c|c|}
    \hline
    $d$ & Box size $N_1\times \cdots \times N_d$ \\
    \hline
    $1$ & $100$ \\
    \hline
    $2$ & $60\times 60$ \\
    \hline
    $3$ & $30\times 30\times 30$ \\
    \hline
    $4$ & $20\times 20\times 20\times 20$ \\
    \hline
    $5$ & $10\times 10\times 10\times 10\times 10$ \\
    \hline
    \end{tabular}
    \caption{Chosen box size as a function of the number of dimensions $d$.}
    \label{tab:box-sizes}
\end{table}

\begin{figure}[H]
    \centering
    \includegraphics[width=1.0\linewidth]{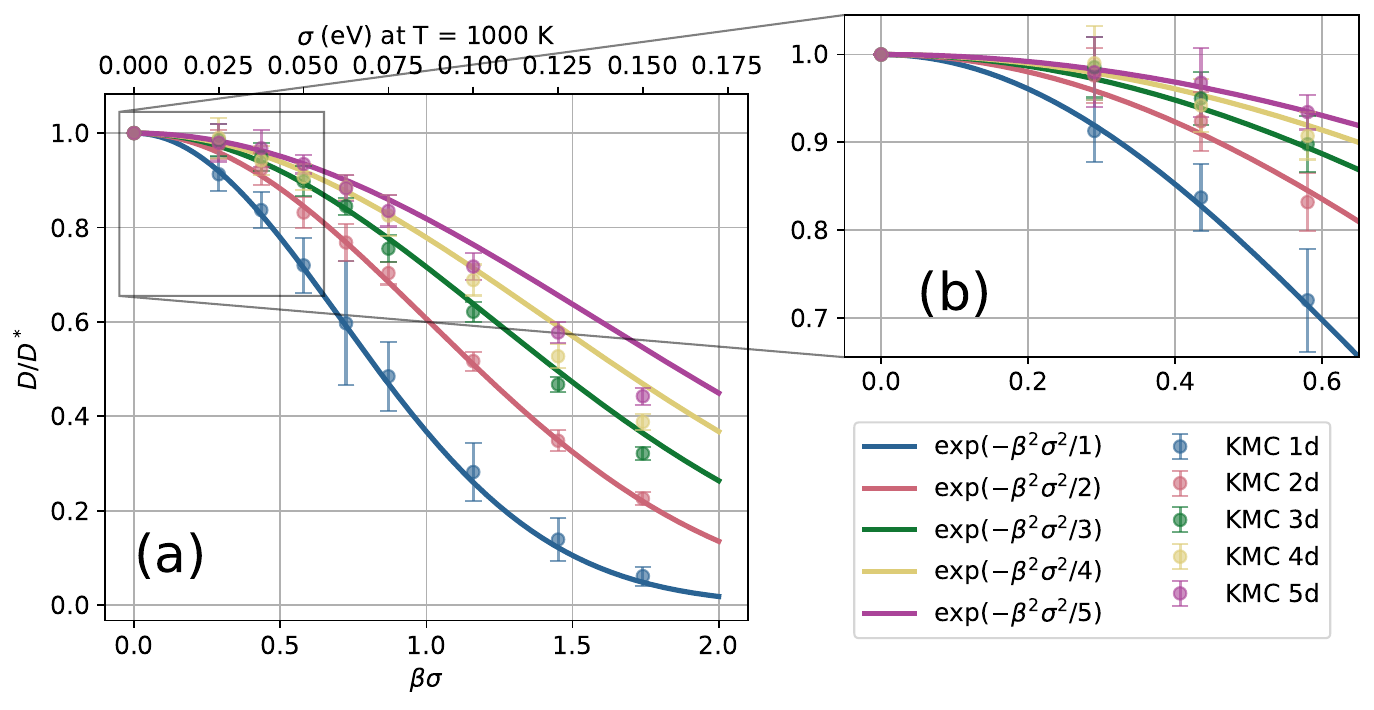}
    \caption{Diffusivity $D$ in terms of the zero-noise diffusivity $D^*$ as a function of solid solution noise $\beta\sigma$, where $\beta = \SI{11.6}{eV^{-1}}$ for the chosen temperature $\SI{1000}{K}$, using KMC and MFPT for a large $\beta\sigma$ range (a) and in the low-noise limit (b).}
    \label{fig:diffusivity}
\end{figure}

For each dimensionality $d$, the diffusivities from KMC and MFPT decrease with the solid solution noise $\sigma$, with each diffusivity decreasing less as the dimensionality $d$ increases (Figure \ref{fig:diffusivity}a). This implies that for a given average potential, the noise always reduces the diffusivity. In the observed cases where the diffusivity increases, the average energy barrier must decrease.

For small $d$, namely $d \leq 3$, the results from KMC and MFPT are in excellent quantitative agreement (Figure \ref{fig:diffusivity}a). However, for large $d$, MFPT overestimates diffusivity. For large $d$ and large $\beta\sigma$, the complexity of the possible paths increases combinatorially and the independent-hop picture underlying the MFPT estimate becomes less controlled. In this case, we hypothesize that transport depends not just on the distribution of individual barriers, but also on how low- and high-barrier transitions are arranged within the network of possible paths. Nevertheless, the MFPT result captures the expected dimensional trend, namely that $D/D^*\to 1$ as $d\to\infty$, consistent with the suppression of energetic disorder when the number of available transitions becomes large.

Interestingly, the approximation remains accurate even when $\beta\sigma$ is not small, despite truncating terms of order $\mathcal{O}(\beta^4\sigma^4)$. Moreover, the low-$\beta\sigma$ truncation agrees with the KMC results substantially better than the full MFPT expression (see Supplementary Material). We interpret this as evidence that the truncated expression acts as an effective trajectory-level approximation, whereas the full single-hop MFPT expression is overly sensitive to the tails of the barrier distribution. At large $\beta\sigma$, rare transitions can strongly bias single-hop averages, either by trapping the particle in superbasins or leading to percolation through the network. Note that in this work, energies are sampled from uncorrelated distributions. These effects likely explain why the truncated expression provides a better description of the ergodic KMC dynamics than the full MFPT expression. 


Furthermore, we have neglected kinetic correlations in hops, which are known to be important for quantifying diffusion in alloys, often significantly reducing diffusivity \cite{Allnatt01102016}. These kinetic correlations have been well-addressed in the literature, including calculation of correlation factors of both vacancies and alloying types in non-dilute alloys \cite{PhysRevB.4.1111}, the generalization of correlation factors for high tracer concentration \cite{SATO19851361}, a mean-field theory approach to the calculation of the full Onsager matrix \cite{Barbe11042006}, and a variational approach for computing transport coefficients \cite{PhysRevLett.121.235901}. This limitation is expected to be most important for vacancy-mediated substitutional diffusion, where the diffusing species remains dynamically coupled to the vacancy after each exchange and successive hops are therefore intrinsically correlated. By contrast, dilute interstitial diffusion is more naturally described as a single walker moving through a network of interstitial sites. Although correlations may still arise from spatial heterogeneity in the local chemical environment, these correlations are not accompanied by the same vacancy-mediated back-correlation mechanism. Consequently, we expect the present independent-hop treatment to be better suited to interstitial diffusion than to vacancy-mediated substitutional diffusion. We expect that the introduction of a spatial correlation length, similar to work done by Banerjee et al. in $1d$ \cite{banerjee2014diffusion}, would be particularly useful to addressing this. However, mimicking this work in more than one dimension is likely not trivial. As such, we save this for a future work.

The model and corresponding expression proposed in this work are valid only in the limit where harmonic transition state theory is valid. For high temperatures, where vibrational entropy might depend on temperature, full transition state theory or its extensions \cite{10.1063/1.4997571}\cite{10.1063/1.461221} should be used, potentially yielding non-Arrhenius expressions for hopping rates\cite{smirnova2020atomistic}.

We have additionally assumed that the energy barriers can be seen as a normal random variable, which might not be true in systems that are not perfectly random. This limitation is particularly relevant in systems where the diffusivity rapidly varies, such as near phase transition boundaries \cite{vattulainen1997non}, non-homogeneous media in which effective diffusivity can strongly vary over space and/or time \cite{pacheco2024langevin, massignan2014nonergodic}, in systems that have a sufficiently strong tendency to chemically order \cite{singh2015atomic}, or in systems where diffusion is otherwise non-Gaussian \cite{luo2019quenched, luo2018non, postnikov2020brownian, pacheco2021convergence}.

We have also implicitly assumed that our system is isotropic, with constant temperature and constant volume. This is particularly relevant for systems under stress, for example. This problem is easily amenable for systems under isotropic stress, where the hopping rates in the solution are:

\begin{equation}
    \begin{aligned}
        \lambda &= \nu\exp(-\beta\Delta H)\\
                &= \nu\exp\left(-\beta(\Delta E + p\Delta V)\right)
    \end{aligned}
\end{equation}

where $\Delta H$ is the enthalpy barrier of a hop, $\Delta V$ is the activation volume of a hop, and $p = \frac{1}{3}\text{Tr}(\boldsymbol{\sigma})$ is the hydrostatic pressure of some stress tensor $\boldsymbol{\sigma}$. Then, if we assume that $\Delta V\sim \mathcal{N}(\mu_V, 2\sigma_V^2)$ is normally distributed, similar to Eq.~\eqref{eq:e-normal-dist}:

\begin{equation}
    \Delta H = \Delta E + p\Delta V \sim \mathcal{N}(\mu + p\mu_V, 2\sigma^2 + 2p^2\sigma_V^2)
\end{equation}

which can then be used in the expression for $\lambda$ in Eq.~\eqref{eq:e-rates}. The case for systems under anisotropic stress, or anisotropic conditions in general, is more nuanced. For example, for a uniaxially stressed system, it is likely more advantageous to define a diffusivity along the stressed axis $D^\parallel$, and a diffusivity along the perpendicular plane $D^\perp$. In this case, our analysis still applies, but using our expression in one dimension for $D^\parallel$ and in two dimensions for $D^\perp$. In the more general case, it is likely advantageous to decompose the stress into principle directions, and perform similar analyses along those axes. However, whether these quantities are of interest, or if an effective isotropic diffusivity is of greater interest, is likely cumbersome and problem-dependent, and is saved for a future work.

This analysis is also for dilute walkers, rather than concentrated ones. In an austenitic steel, for example, this would mean that we can calculate the diffusivity of dilute impurities (such as vacancies, self-interstitials, hydrogen, etc.), but not for the primary alloying elements (iron, nickel, chromium, molybdenum, etc.). However, since diffusion of these primary elements is facilitated by point defects, the energy distribution for the defect-element exchange will shed light on the diffusivity of each element.
Regardless, a more complete theory is needed to compute important transport coefficients, including the full Onsager matrix \cite{ARAllnatt_1982} and interdiffusion coefficients \cite{van2005first}. Recent work has emphasized that activation-like escape times in interacting diffusive systems can acquire density- and interaction-dependent effective barriers \cite{kumar2024arrhenius}, further highlighting that the dilute-walker limit considered here is not expected to directly capture concentrated many-body transport.

Furthermore, we note that the present treatment reduces the diffusion of the walker to a single MFPT. This is the natural quantity entering a coarse-grained long-time diffusivity when hops are treated as renewal events with finite mean residence times. However, the MFPT does not fully characterize the first-passage process. In particular, the full first-passage-time distribution can contain short-time contributions from unusually direct or low-barrier trajectories that are not resolved by the mean alone \cite{godec2016universal, grebenkov2018strong}, which is potentially relevant in higher dimensions (i.e. when the number of explored paths is large), and might be responsible for the inconsistency of our treatment for $d \in \{4, 5\}$ at high $\beta\sigma$ values. We expect this to be problematic in cases where the walker is not immortal and does not see the full distribution of barriers throughout its lifetime \cite{ma2020strong}, e.g. a vacancy diffusing in close proximity to a sink like a grain boundary. A more complete treatment would therefore propagate the full distribution of single-hop first-passage times through the random-walk model, rather than retaining only its first moment. This provides a natural direction for future research, particularly in regimes where disorder produces broad or strongly non-exponential residence-time distributions.

We also note that the independent-rate assumption in the analytical treatment is not exact for the microscopic description where we independently sample initial and saddle point energies. In the KMC simulations, all rates leaving a site share the same initial-site energy \(E_i^\circ\), producing local correlations among the outgoing rates. The analytical model instead treats the post-renormalization directional rates as independent effective channels for the walker. The comparison with KMC is therefore an important part of the validation: the KMC simulations retain these local common-site correlations, while the analytical closure does not. The agreement for \(d\leq 3\) suggests that these correlations are not the leading source of error in the regime where the expression is quantitatively accurate. The larger deviations observed for higher \(d\) and larger \(\beta\sigma\) may reflect the increasing importance of such residual correlations, together with network-connectivity and rare-path effects.

Lastly, our results are derived for simple cubic lattices, and generalizing to other lattice structures might not be trivial. We hypothesize that a similar approach can be utilized for non-cubic structures, and that one can derive an effective dimensionality $d$ from the geometry of said lattice. This would be of particular use to, for example, studying solute diffusion on different faceted surfaces of a nanoparticle, in which each diffusion graph is both regular (that is, has constant degree/coordination number) and two-dimensional, or studying point defect diffusion in bulk bcc or fcc alloys of interest.


In summary, in this work, we have developed an analytical expression for the diffusivity of a dilute walker in a noisy $d$-dimensional solid solution by calculating the mean first passage time (MFPT) as a function of the potential roughness. We then performed kinetic Monte Carlo simulations to calculate the diffusivity as a function of the solid solution noise, which are in excellent agreement with the analytical expression for $d\leq 3$. We observe that the noise always decreases the diffusivity for a given mean activation energy, i.e., if modifying the potential energy surface adding alloying elements increases the diffusivity is because the mean energy barrier decreases. Thus, the theory isolates a roughness-only contribution to sluggish diffusion: it specifies the reference state with respect to which barrier noise is slowing. This result should not be interpreted as a complete theory of sluggish diffusion in concentrated alloys, where species-dependent barriers, chemical bias, vacancy correlations, defect concentrations, and percolating transport pathways may also contribute. We also see that an increase in dimensionality increases the diffusivity, which relates to extra percolation pathways that enhance walker migration. This work opens the possibility of calculating transport coefficients in concentrated random fields without the need of lengthy atomistic simulations, critical to understand the evolution of materials out of equilibrium.
\section*{Data Availability}

The data used in and generated by this work are available from the corresponding authors upon request.
\section*{Acknowledgements}

Authors acknowledge support from the U.S. Department of Energy, Office of Basic Energy Sciences, Materials Science and Engineering Division under Award No. DE-SC0022980.

Additionally, this material is based on work supported by the National Science Foundation under Grant Nos. MRI\# 2024205, MRI\# 1725573, and CRI\# 2010270 for allotment of compute time on the Clemson University Palmetto Cluster.
\section*{Disclaimer}

Any opinions, findings, and conclusions or recommendations expressed in this material are those of the author(s) and do not necessarily reflect the views of the Department of Energy.

\bibliography{bibfile.bib}

@article{10.1063/1.5086746,
    author = {Kadkhodaei, S. and van de Walle, A.},
    title = "{A simple local expression for the prefactor in transition state theory}",
    journal = {The Journal of Chemical Physics},
    volume = {150},
    number = {14},
    pages = {144105},
    year = {2019},
    month = {04},
    issn = {0021-9606},
    doi = {10.1063/1.5086746},
    url = {https://doi.org/10.1063/1.5086746},
    eprint = {https://pubs.aip.org/aip/jcp/article-pdf/doi/10.1063/1.5086746/15558013/144105\_1\_online.pdf},
}

@article{VINEYARD1957121,
title = {Frequency factors and isotope effects in solid state rate processes},
journal = {Journal of Physics and Chemistry of Solids},
volume = {3},
number = {1},
pages = {121-127},
year = {1957},
issn = {0022-3697},
doi = {https://doi.org/10.1016/0022-3697(57)90059-8},
url = {https://www.sciencedirect.com/science/article/pii/0022369757900598},
author = {George H. Vineyard}
}

@article{10.1063/1.1323224,
    author = {Henkelman, Graeme and Jónsson, Hannes},
    title = "{Improved tangent estimate in the nudged elastic band method for finding minimum energy paths and saddle points}",
    journal = {The Journal of Chemical Physics},
    volume = {113},
    number = {22},
    pages = {9978-9985},
    year = {2000},
    month = {12},
    issn = {0021-9606},
    doi = {10.1063/1.1323224},
    url = {https://doi.org/10.1063/1.1323224},
    eprint = {https://pubs.aip.org/aip/jcp/article-pdf/113/22/9978/19260728/9978\_1\_online.pdf},
}

@article{10.1063/1.1329672,
    author = {Henkelman, Graeme and Uberuaga, Blas P. and Jónsson, Hannes},
    title = "{A climbing image nudged elastic band method for finding saddle points and minimum energy paths}",
    journal = {The Journal of Chemical Physics},
    volume = {113},
    number = {22},
    pages = {9901-9904},
    year = {2000},
    month = {12},
    issn = {0021-9606},
    doi = {10.1063/1.1329672},
    url = {https://doi.org/10.1063/1.1329672},
    eprint = {https://pubs.aip.org/aip/jcp/article-pdf/113/22/9901/19259681/9901\_1\_online.pdf},
}

@article{Nakano2008ASP,
  title={A space-time-ensemble parallel nudged elastic band algorithm for molecular kinetics simulation},
  author={Aiichiro Nakano},
  journal={Comput. Phys. Commun.},
  year={2008},
  volume={178},
  pages={280-289},
  url={https://api.semanticscholar.org/CorpusID:1745801}
}

@article{MARAS201613,
title = {Global transition path search for dislocation formation in {Ge} on {Si}(001)},
journal = {Computer Physics Communications},
volume = {205},
pages = {13-21},
year = {2016},
issn = {0010-4655},
doi = {https://doi.org/10.1016/j.cpc.2016.04.001},
url = {https://www.sciencedirect.com/science/article/pii/S0010465516300893},
author = {E. Maras and O. Trushin and A. Stukowski and T. Ala-Nissila and H. Jónsson}
}

@Article{LAMMPS,
  author = "A. P. Thompson and H. M. Aktulga and R. Berger and 
     D. S. Bolintineanu and W. M. Brown and P. S. Crozier and
     P. J. in 't Veld and A. Kohlmeyer and S. G. Moore and T. D. Nguyen and
     R. Shan and M. J. Stevens and J. Tranchida and C. Trott and S. J. Plimpton",
  title = "{LAMMPS} - a flexible simulation tool for
     particle-based materials modeling at the 
     atomic, meso, and continuum scales",
  journal = "Comp. Phys. Comm.",
  volume =  "271",
  pages =   "108171",
  year =    "2022",
  doi = "10.1016/j.cpc.2021.108171"
}

@Article{Case2023,
author={Case, David A.
and Aktulga, Hasan Metin
and Belfon, Kellon
and Cerutti, David S.
and Cisneros, G. Andr{\'e}s
and Cruzeiro, Vin{\'i}cius Wilian D.
and Forouzesh, Negin
and Giese, Timothy J.
and G{\"o}tz, Andreas W.
and Gohlke, Holger
and Izadi, Saeed
and Kasavajhala, Koushik
and Kaymak, Mehmet C.
and King, Edward
and Kurtzman, Tom
and Lee, Tai-Sung
and Li, Pengfei
and Liu, Jian
and Luchko, Tyler
and Luo, Ray
and Manathunga, Madushanka
and Machado, Matias R.
and Nguyen, Hai Minh
and O'Hearn, Kurt A.
and Onufriev, Alexey V.
and Pan, Feng
and Pantano, Sergio
and Qi, Ruxi
and Rahnamoun, Ali
and Risheh, Ali
and Schott-Verdugo, Stephan
and Shajan, Akhil
and Swails, Jason
and Wang, Junmei
and Wei, Haixin
and Wu, Xiongwu
and Wu, Yongxian
and Zhang, Shi
and Zhao, Shiji
and Zhu, Qiang
and Cheatham III, Thomas E.
and Roe, Daniel R.
and Roitberg, Adrian
and Simmerling, Carlos
and York, Darrin M.
and Nagan, Maria C.
and Merz Jr., Kenneth M.},
title={{AmberTools}},
journal={Journal of Chemical Information and Modeling},
year={2023},
month={Oct},
day={23},
publisher={American Chemical Society},
volume={63},
number={20},
pages={6183-6191},
issn={1549-9596},
doi={10.1021/acs.jcim.3c01153},
url={https://doi.org/10.1021/acs.jcim.3c01153}
}

@ARTICLE{Brooks2009-kz,
  title     = "{CHARMM}: the biomolecular simulation program",
  author    = "Brooks, B R and Brooks, 3rd, C L and Mackerell, Jr, A D and
               Nilsson, L and Petrella, R J and Roux, B and Won, Y and
               Archontis, G and Bartels, C and Boresch, S and Caflisch, A and
               Caves, L and Cui, Q and Dinner, A R and Feig, M and Fischer, S
               and Gao, J and Hodoscek, M and Im, W and Kuczera, K and
               Lazaridis, T and Ma, J and Ovchinnikov, V and Paci, E and
               Pastor, R W and Post, C B and Pu, J Z and Schaefer, M and Tidor,
               B and Venable, R M and Woodcock, H L and Wu, X and Yang, W and
               York, D M and Karplus, M",
  journal   = "J. Comput. Chem.",
  publisher = "Wiley",
  volume    =  30,
  number    =  10,
  pages     = "1545--1614",
  month     =  jul,
  year      =  2009,
  copyright = "http://onlinelibrary.wiley.com/termsAndConditions\#vor",
  language  = "en"
}

@article{10.1063/1.1627754,
    author = {Chu, Jhih-Wei and Trout, Bernhardt L. and Brooks, Bernard R.},
    title = "{A super-linear minimization scheme for the nudged elastic band method}",
    journal = {The Journal of Chemical Physics},
    volume = {119},
    number = {24},
    pages = {12708-12717},
    year = {2003},
    month = {12},
    issn = {0021-9606},
    doi = {10.1063/1.1627754},
    url = {https://doi.org/10.1063/1.1627754},
    eprint = {https://pubs.aip.org/aip/jcp/article-pdf/119/24/12708/19253724/12708\_1\_online.pdf},
}

@article{PhysRevB.47.558,
  title = {Ab initio molecular dynamics for liquid metals},
  author = {Kresse, G. and Hafner, J.},
  journal = {Phys. Rev. B},
  volume = {47},
  issue = {1},
  pages = {558--561},
  numpages = {0},
  year = {1993},
  month = {Jan},
  publisher = {American Physical Society},
  doi = {10.1103/PhysRevB.47.558},
  url = {https://link.aps.org/doi/10.1103/PhysRevB.47.558}
}

@article{KRESSE199615,
title = {Efficiency of ab-initio total energy calculations for metals and semiconductors using a plane-wave basis set},
journal = {Computational Materials Science},
volume = {6},
number = {1},
pages = {15-50},
year = {1996},
issn = {0927-0256},
doi = {https://doi.org/10.1016/0927-0256(96)00008-0},
url = {https://www.sciencedirect.com/science/article/pii/0927025696000080},
author = {G. Kresse and J. Furthmüller}
}

@article{PhysRevB.54.11169,
  title = {Efficient iterative schemes for ab initio total-energy calculations using a plane-wave basis set},
  author = {Kresse, G. and Furthm\"uller, J.},
  journal = {Phys. Rev. B},
  volume = {54},
  issue = {16},
  pages = {11169--11186},
  numpages = {0},
  year = {1996},
  month = {Oct},
  publisher = {American Physical Society},
  doi = {10.1103/PhysRevB.54.11169},
  url = {https://link.aps.org/doi/10.1103/PhysRevB.54.11169}
}

@article{Zwanzig1988,
  title = {Diffusion in a rough potential.},
  volume = {85},
  ISSN = {1091-6490},
  url = {http://dx.doi.org/10.1073/pnas.85.7.2029},
  DOI = {10.1073/pnas.85.7.2029},
  number = {7},
  journal = {Proceedings of the National Academy of Sciences},
  publisher = {Proceedings of the National Academy of Sciences},
  author = {Zwanzig,  R},
  year = {1988},
  month = apr,
  pages = {2029–2030}
}

@article{nastar20121,
  title={{1.18-Radiation-induced segregation}},
  author={Nastar, M and Soisson, F},
  journal={Comprehensive nuclear materials},
  volume={1},
  pages={471--496},
  year={2012},
  publisher={Elsevier Oxford}
}

@article{wiedersich1979theory,
  title={A theory of radiation-induced segregation in concentrated alloys},
  author={Wiedersich, H and Okamoto, PR and Lam, Nghi Q},
  journal={Journal of Nuclear Materials},
  volume={83},
  number={1},
  pages={98--108},
  year={1979},
  publisher={Elsevier}
}

@article{rezwan2022effect,
  title={Effect of concurrent grain growth on radiation-induced segregation in nanocrystalline {Fe-Cr-Ni} alloys},
  author={Rezwan, Aashique A and Schwen, Daniel and Zhang, Yongfeng},
  journal={Journal of Nuclear Materials},
  volume={563},
  pages={153614},
  year={2022},
  publisher={Elsevier}
}

@article{SCHEINER20092898,
title = {Finite {Volume} model for diffusion- and activation-controlled pitting corrosion of stainless steel},
journal = {Computer Methods in Applied Mechanics and Engineering},
volume = {198},
number = {37},
pages = {2898-2910},
year = {2009},
issn = {0045-7825},
doi = {https://doi.org/10.1016/j.cma.2009.04.012},
url = {https://www.sciencedirect.com/science/article/pii/S0045782509001790},
author = {Stefan Scheiner and Christian Hellmich}
}

@article{jafarzadeh2019computational,
  title={Computational modeling of pitting corrosion},
  author={Jafarzadeh, Siavash and Chen, Ziguang and Bobaru, Florin},
  journal={Corrosion reviews},
  volume={37},
  number={5},
  pages={419--439},
  year={2019},
  publisher={De Gruyter}
}

@article{CAHN19974397,
title = {A phase-field model for diffusion-induced grain-boundary motion},
journal = {Acta Materialia},
volume = {45},
number = {10},
pages = {4397-4413},
year = {1997},
issn = {1359-6454},
doi = {https://doi.org/10.1016/S1359-6454(97)00074-8},
url = {https://www.sciencedirect.com/science/article/pii/S1359645497000748},
author = {J.W. Cahn and P. Fife and O. Penrose}
}

@article{WANG2006953,
title = {Computer modeling and simulation of solid-state sintering: {A} phase field approach},
journal = {Acta Materialia},
volume = {54},
number = {4},
pages = {953-961},
year = {2006},
issn = {1359-6454},
doi = {https://doi.org/10.1016/j.actamat.2005.10.032},
url = {https://www.sciencedirect.com/science/article/pii/S135964540500635X},
author = {Yu U. Wang}
}

@article{PIOCHAUD2016249,
title = {Atomic-based phase-field method for the modeling of radiation induced segregation in {Fe–Cr}},
journal = {Computational Materials Science},
volume = {122},
pages = {249-262},
year = {2016},
issn = {0927-0256},
doi = {https://doi.org/10.1016/j.commatsci.2016.05.021},
url = {https://www.sciencedirect.com/science/article/pii/S0927025616302452},
author = {J.B. Piochaud and M. Nastar and F. Soisson and L. Thuinet and A. Legris}
}

@article{SENNINGER20161,
title = {Modeling radiation induced segregation in iron–chromium alloys},
journal = {Acta Materialia},
volume = {103},
pages = {1-11},
year = {2016},
issn = {1359-6454},
doi = {https://doi.org/10.1016/j.actamat.2015.09.058},
url = {https://www.sciencedirect.com/science/article/pii/S1359645415007478},
author = {Oriane Senninger and Frédéric Soisson and Enrique Martínez and Maylise Nastar and Chu-Chun Fu and Yves Bréchet}
}

@article{tsai2013sluggish,
  title={Sluggish diffusion in {Co-Cr-Fe-Mn-Ni} high-entropy alloys},
  author={Tsai, K-Y and Tsai, M-H and Yeh, J-W},
  journal={Acta Materialia},
  volume={61},
  number={13},
  pages={4887--4897},
  year={2013},
  publisher={Elsevier}
}

@article{divinski2018mystery,
  title={A mystery of ``sluggish diffusion" in high-entropy alloys: the truth or a myth?},
  author={Divinski, Sergiy V and Pokoev, Alexander V and Esakkiraja, Neelamegan and Paul, Aloke},
  journal={Diffusion foundations},
  volume={17},
  pages={69--104},
  year={2018},
  publisher={Trans Tech Publ}
}

@article{dkabrowa2019demystifying,
  title={Demystifying the sluggish diffusion effect in high entropy alloys},
  author={Dabrowa, Juliusz and Zajusz, Marek and Kucza, Witold and Cie{\'s}lak, Grzegorz and Berent, Katarzyna and Czeppe, Tomasz and Kulik, Tadeusz and Danielewski, Marek},
  journal={Journal of Alloys and Compounds},
  volume={783},
  pages={193--207},
  year={2019},
  publisher={Elsevier}
}

@article{doi:10.1080/01418617908239293,
author = {A. Atkinson and R. I. Taylor},
title = {The diffusion of {Ni} in the bulk and along dislocations in {NiO} single crystals},
journal = {Philosophical Magazine A},
volume = {39},
number = {5},
pages = {581--595},
year = {1979},
publisher = {Taylor \& Francis},
doi = {10.1080/01418617908239293},


URL = { 
    
        https://doi.org/10.1080/01418617908239293
    
    

},
eprint = { 
    
        https://doi.org/10.1080/01418617908239293
    
    

}
}

@article{wang2006single,
  title={Single molecule measurements of repressor protein {1D} diffusion on {DNA}},
  author={Wang, YM and Austin, Robert H and Cox, Edward C},
  journal={Physical review letters},
  volume={97},
  number={4},
  pages={048302},
  year={2006},
  publisher={APS}
}

@article{gorman2010visualizing,
  title={Visualizing one-dimensional diffusion of eukaryotic {DNA} repair factors along a chromatin lattice},
  author={Gorman, Jason and Plys, Aaron J and Visnapuu, Mari-Liis and Alani, Eric and Greene, Eric C},
  journal={Nature structural \& molecular biology},
  volume={17},
  number={8},
  pages={932--938},
  year={2010},
  publisher={Nature Publishing Group US New York}
}

@article{gorman2008visualizing,
  title={Visualizing one-dimensional diffusion of proteins along {DNA}},
  author={Gorman, Jason and Greene, Eric C},
  journal={Nature structural \& molecular biology},
  volume={15},
  number={8},
  pages={768--774},
  year={2008},
  publisher={Nature Publishing Group US New York}
}

@article{legros2008observation,
  title={Observation of giant diffusivity along dislocation cores},
  author={Legros, Marc and Dehm, Gerhard and Arzt, Eduard and Balk, T John},
  journal={Science},
  volume={319},
  number={5870},
  pages={1646--1649},
  year={2008},
  publisher={American Association for the Advancement of Science}
}

@article{mishin1997grain,
  title={Grain boundary diffusion: fundamentals to recent developments},
  author={Mishin, Yu and Herzig, Chr and Bernardini, J and Gust, W},
  journal={International materials reviews},
  volume={42},
  number={4},
  pages={155--178},
  year={1997},
  publisher={SAGE Publications Sage UK: London, England}
}

@article{peterson1983grain,
  title={Grain-boundary diffusion in metals},
  author={Peterson, NL},
  journal={International metals reviews},
  volume={28},
  number={1},
  pages={65--91},
  year={1983},
  publisher={SAGE Publications Sage UK: London, England}
}

@article{vaidya2018bulk,
  title={Bulk tracer diffusion in {CoCrFeNi} and {CoCrFeMnNi} high entropy alloys},
  author={Vaidya, M and Pradeep, KG and Murty, BS and Wilde, G and Divinski, SV},
  journal={Acta Materialia},
  volume={146},
  pages={211--224},
  year={2018},
  publisher={Elsevier}
}

@article{beke2016diffusion,
  title={On the diffusion in high-entropy alloys},
  author={Beke, DL and Erd{\'e}lyi, G},
  journal={Materials Letters},
  volume={164},
  pages={111--113},
  year={2016},
  publisher={Elsevier}
}

@article{vaidya2016ni,
  title={Ni tracer diffusion in {CoCrFeNi} and {CoCrFeMnNi} high entropy alloys},
  author={Vaidya, Mayur and Trubel, Simon and Murty, BS and Wilde, Gerhard and Divinski, Sergiy V},
  journal={Journal of Alloys and Compounds},
  volume={688},
  pages={994--1001},
  year={2016},
  publisher={Elsevier}
}

@article{RevModPhys.15.1,
  title = {{Stochastic Problems in Physics and Astronomy}},
  author = {Chandrasekhar, S.},
  journal = {Rev. Mod. Phys.},
  volume = {15},
  issue = {1},
  pages = {1--89},
  numpages = {0},
  year = {1943},
  month = {Jan},
  publisher = {American Physical Society},
  doi = {10.1103/RevModPhys.15.1},
  url = {https://link.aps.org/doi/10.1103/RevModPhys.15.1}
}

@article{BORTZ197510,
title = {A new algorithm for {Monte Carlo} simulation of {Ising} spin systems},
journal = {Journal of Computational Physics},
volume = {17},
number = {1},
pages = {10-18},
year = {1975},
issn = {0021-9991},
doi = {https://doi.org/10.1016/0021-9991(75)90060-1},
url = {https://www.sciencedirect.com/science/article/pii/0021999175900601},
author = {A.B. Bortz and M.H. Kalos and J.L. Lebowitz}
}

@article{vattulainen1997non,
  title={{Non-Arrhenius} behavior of surface diffusion near a phase transition boundary},
  author={Vattulainen, I and Merikoski, J and Ala-Nissila, T and Ying, SC},
  journal={Physical review letters},
  volume={79},
  number={2},
  pages={257},
  year={1997},
  publisher={APS}
}

@article{singh2015atomic,
  title={Atomic short-range order and incipient long-range order in high-entropy alloys},
  author={Singh, Prashant and Smirnov, Andrei V and Johnson, Duane D},
  journal={Physical Review B},
  volume={91},
  number={22},
  pages={224204},
  year={2015},
  publisher={APS}
}

@article{smirnova2020atomistic,
  title={Atomistic description of self-diffusion in molybdenum: A comparative theoretical study of non-{Arrhenius} behavior},
  author={Smirnova, Daria and Starikov, Sergei and Leines, Grisell D{\'\i}az and Liang, Yanyan and Wang, Ning and Popov, Maxim N and Abrikosov, Igor A and Sangiovanni, Davide G and Drautz, Ralf and Mrovec, Matous},
  journal={Physical Review Materials},
  volume={4},
  number={1},
  pages={013605},
  year={2020},
  publisher={APS}
}

@article{htst,
  title={The activated complex and the absolute rate of chemical reactions.},
  author={Eyring, Henry},
  journal={Chemical Reviews},
  volume={17},
  number={1},
  pages={65-77},
  year={1935},
  publisher={ACS}
}

@article{10.1063/1.4997571,
    author = {Revuelta, F. and Craven, Galen T. and Bartsch, Thomas and Borondo, F. and Benito, R. M. and Hernandez, Rigoberto},
    title = {Transition state theory for activated systems with driven anharmonic barriers},
    journal = {The Journal of Chemical Physics},
    volume = {147},
    number = {7},
    pages = {074104},
    year = {2017},
    month = {08},
    issn = {0021-9606},
    doi = {10.1063/1.4997571},
    url = {https://doi.org/10.1063/1.4997571},
    eprint = {https://pubs.aip.org/aip/jcp/article-pdf/doi/10.1063/1.4997571/15530768/074104\_1\_online.pdf},
}

@article{10.1063/1.461221,
    author = {Gonzalez‐Lafont, Angels and Truong, Thanh N. and Truhlar, Donald G.},
    title = {Interpolated variational transition‐state theory: Practical methods for estimating variational transition‐state properties and tunneling contributions to chemical reaction rates from electronic structure calculations},
    journal = {The Journal of Chemical Physics},
    volume = {95},
    number = {12},
    pages = {8875-8894},
    year = {1991},
    month = {12},
    issn = {0021-9606},
    doi = {10.1063/1.461221},
    url = {https://doi.org/10.1063/1.461221},
    eprint = {https://pubs.aip.org/aip/jcp/article-pdf/95/12/8875/18996914/8875\_1\_online.pdf},
}

@article{ternary-eam,
author = {Zhou, Xiaowang W. and Foster, Michael E. and Sills, Ryan B.},
title = {{An Fe-Ni-Cr embedded atom method potential for austenitic and ferritic systems}},
journal = {Journal of Computational Chemistry},
volume = {39},
number = {29},
pages = {2420-2431},
doi = {https://doi.org/10.1002/jcc.25573},
url = {https://onlinelibrary.wiley.com/doi/abs/10.1002/jcc.25573},
eprint = {https://onlinelibrary.wiley.com/doi/pdf/10.1002/jcc.25573},
year = {2018}
}

@article{Allnatt01102016,
author = {A. R. Allnatt and T. R. Paul and I. V. Belova and G. E. Murch},
title = {A high accuracy diffusion kinetics formalism for random multicomponent alloys: application to high entropy alloys},
journal = {Philosophical Magazine},
volume = {96},
number = {28},
pages = {2969--2985},
year = {2016},
publisher = {Taylor \& Francis},
doi = {10.1080/14786435.2016.1219785},


URL = { 
    
        https://doi.org/10.1080/14786435.2016.1219785
    
    

},
eprint = { 
    
        https://doi.org/10.1080/14786435.2016.1219785
    
    

}

}

@article{PhysRevB.4.1111,
  title = {Correlation Factors for Diffusion in Nondilute Alloys},
  author = {Manning, John R.},
  journal = {Phys. Rev. B},
  volume = {4},
  issue = {4},
  pages = {1111--1121},
  numpages = {0},
  year = {1971},
  month = {Aug},
  publisher = {American Physical Society},
  doi = {10.1103/PhysRevB.4.1111},
  url = {https://link.aps.org/doi/10.1103/PhysRevB.4.1111}
}

@article{SATO19851361,
title = {Correlation factor in tracer diffusion for high tracer concentrations},
journal = {Journal of Physics and Chemistry of Solids},
volume = {46},
number = {12},
pages = {1361-1370},
year = {1985},
issn = {0022-3697},
doi = {https://doi.org/10.1016/0022-3697(85)90074-5},
url = {https://www.sciencedirect.com/science/article/pii/0022369785900745},
author = {Hiroshi Sato and Takuma Ishikawa and Ryoichi Kikuchi}
}

@article{Barbe11042006,
author = {V. Barbe and M. Nastar},
title = {A self-consistent mean field calculation of the phenomenological coefficients in a multicomponent alloy with high jump frequency ratios},
journal = {Philosophical Magazine},
volume = {86},
number = {11},
pages = {1513--1538},
year = {2006},
publisher = {Taylor \& Francis},
doi = {10.1080/14786430500383575},


URL = { 
    
        https://doi.org/10.1080/14786430500383575
    
    

},
eprint = { 
    
        https://doi.org/10.1080/14786430500383575
    
    

}

}

@article{PhysRevLett.121.235901,
  title = {Variational Principle for Mass Transport},
  author = {Trinkle, Dallas R.},
  journal = {Phys. Rev. Lett.},
  volume = {121},
  issue = {23},
  pages = {235901},
  numpages = {7},
  year = {2018},
  month = {Dec},
  publisher = {American Physical Society},
  doi = {10.1103/PhysRevLett.121.235901},
  url = {https://link.aps.org/doi/10.1103/PhysRevLett.121.235901}
}

@article{ARAllnatt_1982,
doi = {10.1088/0022-3719/15/27/016},
url = {https://doi.org/10.1088/0022-3719/15/27/016},
year = {1982},
month = {sep},
publisher = {},
volume = {15},
number = {27},
pages = {5605},
author = {A R Allnatt},
title = {Einstein and linear response formulae for the phenomenological coefficients for isothermal matter transport in solids},
journal = {Journal of Physics C: Solid State Physics}
}

@article{van2005first,
  title={First principles calculation of the interdiffusion coefficient in binary alloys},
  author={Van der Ven, A and Ceder, G},
  journal={Physical review letters},
  volume={94},
  number={4},
  pages={045901},
  year={2005},
  publisher={APS}
}

@article{seki2015relationship,
  title={Relationship between entropy and diffusion: A statistical mechanical derivation of Rosenfeld expression for a rugged energy landscape},
  author={Seki, Kazuhiko and Bagchi, Biman},
  journal={The Journal of chemical physics},
  volume={143},
  number={19},
  year={2015},
  publisher={AIP Publishing}
}

@article{seki2016anomalous,
  title={Anomalous dimensionality dependence of diffusion in a rugged energy landscape: How pathological is one dimension?},
  author={Seki, Kazuhiko and Bagchi, Kaushik and Bagchi, Biman},
  journal={The Journal of chemical physics},
  volume={144},
  number={19},
  year={2016},
  publisher={AIP Publishing}
}

@article{godec2016universal,
  title={Universal proximity effect in target search kinetics in the few-encounter limit},
  author={Godec, Alja{\v{z}} and Metzler, Ralf},
  journal={Physical Review X},
  volume={6},
  number={4},
  pages={041037},
  year={2016},
  publisher={APS}
}

@article{grebenkov2018strong,
  title={Strong defocusing of molecular reaction times results from an interplay of geometry and reaction control},
  author={Grebenkov, Denis S and Metzler, Ralf and Oshanin, Gleb},
  journal={Communications Chemistry},
  volume={1},
  number={1},
  pages={96},
  year={2018},
  publisher={Nature Publishing Group UK London}
}

@article{ma2020strong,
  title={Strong intracellular signal inactivation produces sharper and more robust signaling from cell membrane to nucleus},
  author={Ma, Jingwei and Do, Myan and Le Gros, Mark A and Peskin, Charles S and Larabell, Carolyn A and Mori, Yoichiro and Isaacson, Samuel A},
  journal={PLoS computational biology},
  volume={16},
  number={11},
  pages={e1008356},
  year={2020},
  publisher={Public Library of Science San Francisco, CA USA}
}

@article{kumar2025speeding,
  title={Speeding up {Brownian} escape via intermediate finite potential barriers},
  author={Kumar, Vishwajeet and Shpielberg, Ohad and Pal, Arnab},
  journal={Chaos: An Interdisciplinary Journal of Nonlinear Science},
  volume={35},
  number={11},
  year={2025},
  publisher={AIP Publishing}
}

@article{kumar2024arrhenius,
  title={Arrhenius law for interacting diffusive systems},
  author={Kumar, Vishwajeet and Pal, Arnab and Shpielberg, Ohad},
  journal={Physical Review E},
  volume={109},
  number={3},
  pages={L032101},
  year={2024},
  publisher={APS}
}

@article{luo2019quenched,
  title={Quenched trap model on the extreme landscape: the rise of sub-diffusion and non-{Gaussian} diffusion},
  author={Luo, Liang and Yi, Ming},
  journal={arXiv preprint arXiv:1906.08294},
  year={2019}
}

@article{luo2018non,
  title={Non-{Gaussian} diffusion in static disordered media},
  author={Luo, Liang and Yi, Ming},
  journal={Physical Review E},
  volume={97},
  number={4},
  pages={042122},
  year={2018},
  publisher={APS}
}

@article{postnikov2020brownian,
  title={Brownian yet non-{Gaussian} diffusion in heterogeneous media: from superstatistics to homogenization},
  author={Postnikov, Eugene B and Chechkin, Aleksei and Sokolov, Igor M},
  journal={New Journal of Physics},
  volume={22},
  number={6},
  pages={063046},
  year={2020},
  publisher={IOP Publishing}
}

@article{pacheco2024langevin,
  title={Langevin equation in heterogeneous landscapes: how to choose the interpretation},
  author={Pacheco-Pozo, Adrian and Balcerek, Micha{\l} and Wy{\l}omanska, Agnieszka and Burnecki, Krzysztof and Sokolov, Igor M and Krapf, Diego},
  journal={Physical Review Letters},
  volume={133},
  number={6},
  pages={067102},
  year={2024},
  publisher={APS}
}

@article{pacheco2021convergence,
  title={Convergence to a {Gaussian} by narrowing of central peak in Brownian yet non-{Gaussian} diffusion in disordered environments},
  author={Pacheco-Pozo, Adrian and Sokolov, Igor M},
  journal={Physical Review Letters},
  volume={127},
  number={12},
  pages={120601},
  year={2021},
  publisher={APS}
}

@article{massignan2014nonergodic,
  title={Nonergodic subdiffusion from {Brownian} motion in an inhomogeneous medium},
  author={Massignan, Pietro and Manzo, Carlo and Torreno-Pina, Juan A and Garc{\'\i}a-Parajo, Maria F and Lewenstein, Maciej and Lapeyre Jr, Gerald J},
  journal={Physical review letters},
  volume={112},
  number={15},
  pages={150603},
  year={2014},
  publisher={APS}
}

@article{xu2023mechanism,
  title={Mechanism of sluggish diffusion under rough energy landscape},
  author={Xu, Biao and Zhang, Jun and Xiong, Yaoxu and Ma, Shihua and Osetsky, Yuri and Zhao, Shijun},
  journal={Cell Reports Physical Science},
  volume={4},
  number={4},
  year={2023},
  publisher={Elsevier}
}

@article{banerjee2014diffusion,
  title={Diffusion on a rugged energy landscape with spatial correlations},
  author={Banerjee, Saikat and Biswas, Rajib and Seki, Kazuhiko and Bagchi, Biman},
  journal={The Journal of chemical physics},
  volume={141},
  number={12},
  year={2014},
  publisher={AIP Publishing}
}

\end{document}